\documentclass[sigconf,10pt,nonacm]{acmart}
\setcopyright{rightsretained}

\usepackage{graphicx} 
\usepackage{amsmath}
\usepackage{array}
\usepackage{booktabs}
\usepackage{enumitem}
\usepackage{xcolor}
\usepackage{subcaption}
\usepackage{cleveref}
\usepackage{tabularx} 
\newcolumntype{Y}[1]{>{\hsize=#1\hsize\raggedright\arraybackslash}X}

\title{Denial of Deadline: Network-Driven Accuracy Collapse in Distributed Inference Pipelines}

\author{Jhonatan Tavori$^{*}$, Gur-Eyal Sela$^{\dagger}$, Ion Stoica$^{\dagger}$, Gil Zussman$^{*}$}
\affiliation{
\institution{$^{*}$Columbia University \qquad $^{\dagger}$UC Berkeley}
\country{}
}

\renewcommand{\shortauthors}{Jhonatan Tavori, Gur-Eyal Sela, Ion Stoica, Gil Zussman}

\begin{abstract}
Inference systems increasingly combine a \textit{fast path} that returns predictions within the application's latency deadline together with a higher-accuracy \textit{slow path} that runs higher-compute methods on stronger, remote hardware, so its results can be returned on time and combined with the fast path predictions. Across several application domains, we abstract this inference architecture as a fast path, a slow path, and a coordination layer with two functions: a router that invokes the slow path and a merger that decides whether to incorporate its returned predictions. In this work, we show that this new coordination layer exposes a new attack surface: shaped workload attacks, e.g., Yo-Yo bursts, can exploit contention at shared resources along the slow path, including cloud autoscaling bottlenecks, to push benign users’ slow-path predictions past their latency deadlines. The merger then discards those predictions, while the fast path continues to return timely outputs. We refer to the resulting loss of slow-path accuracy benefits as \emph{accuracy collapse}.

We demonstrate accuracy collapse in a two-tier edge--cloud multi-object tracking pipeline in autonomous driving. In simulation, approximately 4,000 burst-shaped requests increase benign slow-path p99 latency from 92\,ms to 2\,s. These delays nearly eliminate the benefit of the slow path's cloud inference, reducing object tracking quality by 7.0 HOTA points on average. We further find that accuracy degradation can significantly vary (2.0 - 18.7 HOTA points), depending on the video intervals that are targeted in the attack, and that certain rare classes (e.g., stop signs) in the dataset lose nearly half of their pre-attack prediction accuracy. These results show that workload attacks on shared inference infrastructure can degrade prediction quality without needing either access to model weights or victim data, and motivate research on attacks and defenses for routing, merging, admission control, scheduling, and resource isolation in these emerging inference pipeline architectures.
\end{abstract}

\begin{document}

\maketitle

\section{Introduction}

An increasing number of latency-sensitive inference services combine models or
execution paths with different latency, cost, and capability profiles. More
capable endpoint hardware and compact models allow cameras, mobile devices, and
robots to perform useful inference locally. At the same time, remote
higher-capacity resources can provide higher-quality predictions or more
expensive computation than can be supported locally. Systems therefore combine
a lower-cost, low-latency \emph{fast path} that runs inference locally with a
slower, higher accuracy \emph{slow path} that is invoked selectively to run on
stronger, remote hardware.

This architecture appears in edge–cloud video analytics~\cite{noscope,reducto,dds,glimpse}, routed and cascaded models~\cite{frugalgpt,routellm,hybridllm,tabi}, split and early-exit DNNs~\cite{neurosurgeon,spinn,branchynet,msdnet}, and autonomous systems with remote assistance~\cite{saycan,rt2,pi0,waymo}.

Across these domains, the fast path operates independently, with optional slow-path refinement.
The coordination layer in these
systems can be represented as two logical functions: an inference \emph{router}  decides
when to invoke the slow path, while a \emph{merger} decides whether and how to
use its results. These functions may be explicit  or embedded in a cascade or application policy, and they control when and
how models across this distributed inference system are invoked.
The incremental benefit of the slow path varies across inputs and often decays with delay. Thus, even a semantically correct result may provide little or no benefit if it arrives after the merger’s deadline. Because these pipelines share resources along the slow assistance path, they expose a new attack opportunity.

We study how the trend toward burst-shaped workload attacks \cite{sides2015yo,bremler2017ddos,alcoz2022aggregate,gu2024sync,gu2024grunt} extends to distributed inference. Contention at shared resources, from edge links and gateways to cloud queues and delayed scale-out, can push slow-path results beyond their useful window. The merger then discards otherwise correct assistance, reducing the accuracy realized by the application. We call this failure \emph{accuracy collapse}.

We consider an adversary that submits valid requests and controls only their content and timing. The attacker may choose inputs that are more likely to invoke slow-path processing (e.g., scenes with many or rare objects) and concentrate them into short bursts, inducing contention on shared slow-path resources. Consequently, benign slow-path results miss the merger’s deadline and are discarded, leaving the fast-path prediction as the application output. The attack requires neither model weights nor privileged infrastructure access, yet still induces accuracy collapse for benign users.

This form of accuracy degradation is not captured by endpoint availability alone. Because the fast path continues to return results, the service appears operational even as the benefit of slow-path assistance disappears.
Directly quantifying this loss requires labels or application-specific quality proxies, which may be unavailable during online operation.
As we show, aggregate metrics can obscure the effect further: 
attacks that appear moderate over a complete stream may become catastrophic when they coincide with hard, rare, or safety-critical inputs that depend most on the slow path.

We evaluate this mechanism by simulating the edge-cloud video analytics pipeline deployed on the NSF COSMOS testbed \cite{raychaudhuri2020challenge},  
with a replay of slow-path completion times on an
Argoverse-HD multi-object-tracking pipeline, a latency sensitive perception
pipeline used in safety-critical autonomous driving settings. In our
simulation, burst-shaped attacker traffic raises benign slow-path p99
completion time from roughly 92 ms to over 2 s, far exceeding the
configured 250 ms SLO. When these completion times are replayed against the
tracking pipeline's deadline, delayed cloud predictions remove most of the
accuracy gain from slow-path assistance during attack-active intervals.
The effects vary and include substantial losses for several
low-frequency classes. This demonstrates potency: 
by using burst shaping rather than flat DDoS traffic, the attacker causes disproportionate accuracy harm with the same average attack request rate.

This paper makes the following contributions:
\begin{itemize}

  \item \textbf{A cross-domain systems abstraction.} 
  In Section~\ref{sec:background}, we characterize distributed assisted inference as a fast path, a slow path, and a router--merger coordination layer that determines when remote assistance is invoked and whether its result contributes to the application output.

  \item \textbf{A coordination-layer attack and accuracy evaluation.}
Section~\ref{sec:attack} shows how bursts of otherwise valid requests can exploit contention at shared slow-path resources, including queueing, autoscaling lag, and merge deadlines, to delay benign assistance beyond its useful window. Section~\ref{sec:eval} evaluates this attack and demonstrates aggregate, temporal, and per-class accuracy degradation, which we term \emph{accuracy collapse}.
  \item \textbf{Positioning and research agenda.} 
  Section~\ref{sec-relatedwork} situates the attack within the broader evolution of burst-shaped workload attacks. Section~\ref{sec:discussion} outlines open research directions for resilient distributed inference.

\end{itemize}

\section{Two-System Inference Pipelines}
\label{sec:background}
This section surveys two-system inference pipelines, a distributed architecture that combines a fast local path with a slower remote one. We show how this architecture introduces a coordination layer that exposes a new attack surface.

\begin{figure}[t]
\centering
\includegraphics[width=\columnwidth]{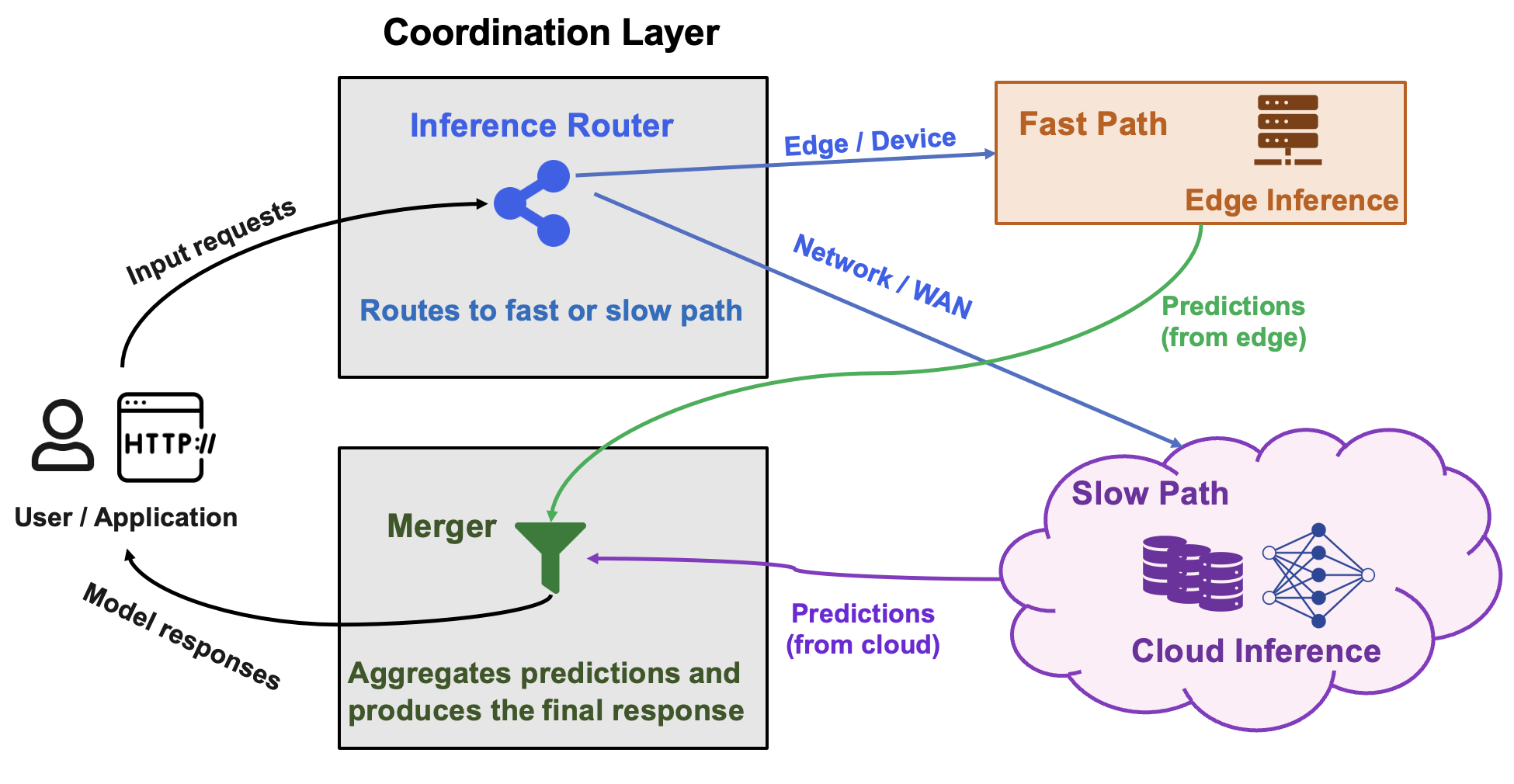}
\caption{A two-system inference pipeline. A \emph{router} decides whether to serve each request on the edge model, the cloud model, or both; the edge and cloud predictions return asynchronously across the network; and a \emph{merger} aggregates them into the final response. The router and merger form the coordination layer.
}
\label{fig:coord}
\end{figure}

\subsection{Why combine fast and slow inference?}

Latency-sensitive edge applications, including video
analytics~\cite{dedelayed,edgeduet}, mobile and wearable
systems~\cite{eaar,glimpse}, and autonomous
robots~\cite{schafhalter2023cloud,asyncvla}, often choose between two execution
paths that trade off prediction quality against latency and resource cost. In
the first, \textit{fast path}, a small, local model can reliably return
predictions within the application's response deadline. In the second,
\textit{slow path}, a larger model runs on remote, higher-capacity datacenter
hardware and may achieve significantly higher accuracy, but its use incurs
additional cost and adds variable network and queuing delay. For example,
Schafhalter et al. report that a high-accuracy, large object detection model
may have an inference latency of 859\,ms on an NVIDIA Orin edge device, but
147\,ms on an A100 datacenter accelerator, while the supporting 5G network exhibits a median RTT of 68\,ms but a p99 of 3.0\,s~\cite{schafhalter2023cloud}.

Since neither execution path alone provides the desired operating point,
recent inference systems increasingly run both in parallel: In the fast path, a
local model returns a prediction for every input, and simultaneously the slow
path is selectively invoked to return higher accuracy predictions to improve
the end-to-end accuracy of the system. The fast path provides a timely baseline
prediction, and the system incorporates a slow-path prediction only if it
returns before a certain freshness deadline when it can still be used by the
system to boost accuracy through merging or input selection.

\begin{table*}[t]
\centering
\footnotesize 
\setlength{\tabcolsep}{3pt} 
\renewcommand{\arraystretch}{1.05}

\caption{Representative two-system inference pipelines across domains.}
\label{tab:currentsystems}

\begin{tabularx}{\textwidth}{@{}
    Y{1.10}
    Y{0.90}
    Y{1.00}
    Y{0.95}
    Y{1.05}
    @{}}
\toprule
\textbf{Use case} &
\textbf{Fast path} &
\textbf{Slow path} &
\textbf{Router} &
\textbf{Merger} \\
\midrule

\textbf{Vehicle and robot perception}%
~\cite{schafhalter2023cloud,emp,carcel,shi2024soar}
&
A small, onboard or on-device model
&
Larger, higher-capacity models
&
Select services, frames, or frame regions to offload
&
Slow path output if on time, otherwise fast path
\\

\textbf{Robot planning and control}%
~\cite{asyncvla,waymo,dualvln,rfst}
&
Onboard controller policy or autonomous driver
&
Remote VLA, semantic planner, or human guidance
&
Periodic or uncertainty-triggered
&
Fuse returned guidance/actions
\\

\textbf{Mobile and edge video analytics}%
~\cite{glimpse,eaar,edgeduet,dedelayed}
&
Small/lightweight detection, tracking
&
Larger detection tracking or prediction correction
&
Select key frames, frame regions, or uncertain inputs
&
Update the tracker or prediction cache
\\

\textbf{Language-model routing} \emph{(related variant)}%
~\cite{applefm,hybridllm,frugalgpt,routellm}
&
SLM, on-device model
&
Frontier LLM model
&
Route by difficulty, quality, privacy, or cost
&
Usually select or summarize to one answer
\\

\bottomrule
\end{tabularx}
\end{table*}

\subsection{Architecture and execution semantics}
\label{sec:arch}

The two-system inference pipeline contains two logical inference paths whose outputs may
contribute to each application result (Figure~\ref{fig:coord}). For an input
$x_t$, the \emph{\textbf{fast path}} runs unconditionally and produces a complete,
usable prediction $f_t$ within the application's end-to-end inference deadline.
When invoked, the optional \emph{\textbf{slow path}} produces a prediction $s_t$ using
higher-capability remote resources. It often provides higher expected accuracy,
but at greater resource cost and with more variable end-to-end latency.
``\textit{Fast}'' and ``\textit{slow}'' denote logical roles, not 
specific machines or 
model speeds.

The \emph{\textbf{router}}\footnote{``Router'' refers to the pipeline's inference-routing component that decides whether to invoke the slow path, not a network router.} decides \emph{whether and where} to invoke the slow path in
addition to the fast path. Its inputs may include $x_t$, the fast-path
prediction or confidence, estimated slow-path benefit, current load, and the
remaining latency budget~\cite{eaar,edgeduet,hybridllm,dedelayed}. Its output
is a dispatch decision linking the slow-path request to the originating input.
Examples include frame selection in edge-cloud video
analytics~\cite{eaar,edgeduet}, a confidence threshold in a model
cascade~\cite{frugalgpt,tabi}, and an uncertainty trigger for remote robot
assistance~\cite{waymo}.

The \emph{\textbf{merger}} determines which available predictions contribute to the
application output. It receives $f_t$ and any returned slow-path output
together with enough provenance and timing information to associate each result
with its source input and assess its age ~\cite{schafhalter2023cloud}.
Depending on the application, it may return fast-path predictions
while incorporating delayed slow-path results into later outputs~\cite{eaar,glimpse}, or wait until the deadline and select the best available result~\cite{schafhalter2023cloud}.
Other systems merge fast- and slow-path predictions, such as cached detections or delayed remote features~\cite{edgeduet,dedelayed}. The merger therefore applies an application-specific freshness policy: stale results may remain useful after temporal propagation, but their accuracy benefit declines as the stream evolves~\cite{glimpse,eaar,dedelayed}.
If no
admissible slow-path result is available by the deadline, the fast-path
prediction remains the output~\cite{schafhalter2023cloud,dedelayed}.
\\
\\
\textbf{Current systems}
The same fast/slow structure recurs across domains that differ in modality,
latency budget, and how the predictions are combined when returned from the
slow path. Table~\ref{tab:currentsystems} surveys representative pipelines: in
each, the fast path supplies a prediction that makes the latency deadline, a
slow path supplies a higher accuracy prediction, and a router and merger coordinate
when the slow path is invoked and how its returned output is handled.

\subsection{The Attack Surface: Shared Resources}
\textbf{Normal operation.}
For every input, the fast path produces a usable prediction by the application’s end-to-end deadline. When invoked, the slow path is intended to arrive before the merger’s deadline and improve the prediction. The fast-path output serves as a \textit{fallback} for the occasional case in which the slow-path result is unavailable or arrives too late.
\\
\textbf{Shared slow-path resources.}
Unlike the typically local fast path, slow-path requests may traverse shared network infrastructure (e.g., network links) or consume shared compute resources (e.g., scaled cloud resources). Consequently, one client’s request can delay assistance for other clients sharing those resources.
\\
\textbf{Attack opportunity.}
By design, the fast-path fallback preserves availability when slow-path assistance is unavailable. An adversary can exploit this by inducing contention that pushes benign assistance past its deadline.
In the following, we define the threat model and present the attack mechanism that exploits this architectural property.

\section{Request-Induced Accuracy Loss}
\label{sec:attack}

We consider a workload-shaping adversary that interacts only through the normal request interface. The adversary aims to collapse the realized value of assistance, rather than necessarily alter the correctness of any individual prediction.

\subsection{Threat Model}
\label{sec:threat-model}

The adversary submits valid inputs as an ordinary user and requires no privileged access. Its requests share network and compute resources with benign requests along the slow path, and it controls their content and timing. In particular, it may select inputs that are likely to trigger slow-path assistance, such as complex video scenes. As we show, its leverage is asymmetric: modest workloads trigger disproportionate delay and accuracy loss. This distinguishes our setting from volumetric DDoS and aligns it with burst-style attacks, such as Yo-Yo attacks~\cite{sides2015yo}, discussed in Section~\ref{sec-relatedwork}. 

The adversary may know the high-level architecture of the system but not its internal parameters such as model weights, exact routing thresholds, or merger rules. It may nevertheless infer routing behavior through probing and use observable signals, including response latency or a returned model identifier, to increase the attack’s potency.

We intentionally assume a weak adversary with no infrastructure control. It cannot inspect or modify other users’ traffic or compromise any component of the inference pipeline. 
Thus, while we do not modify model weights or consider training-time attacks, we induce incorrect application outputs by exploiting contention at shared resources along the assistance path.
This weak threat model highlights that accuracy collapse and more generally the effective performance of the AI pipeline, can result from networking effects even when the models themselves are not compromised.

\subsection{Attack Mechanism}
\label{sec:attack-mechanism}

\begin{figure}[t]
  \centering
  \includegraphics[width=\linewidth]{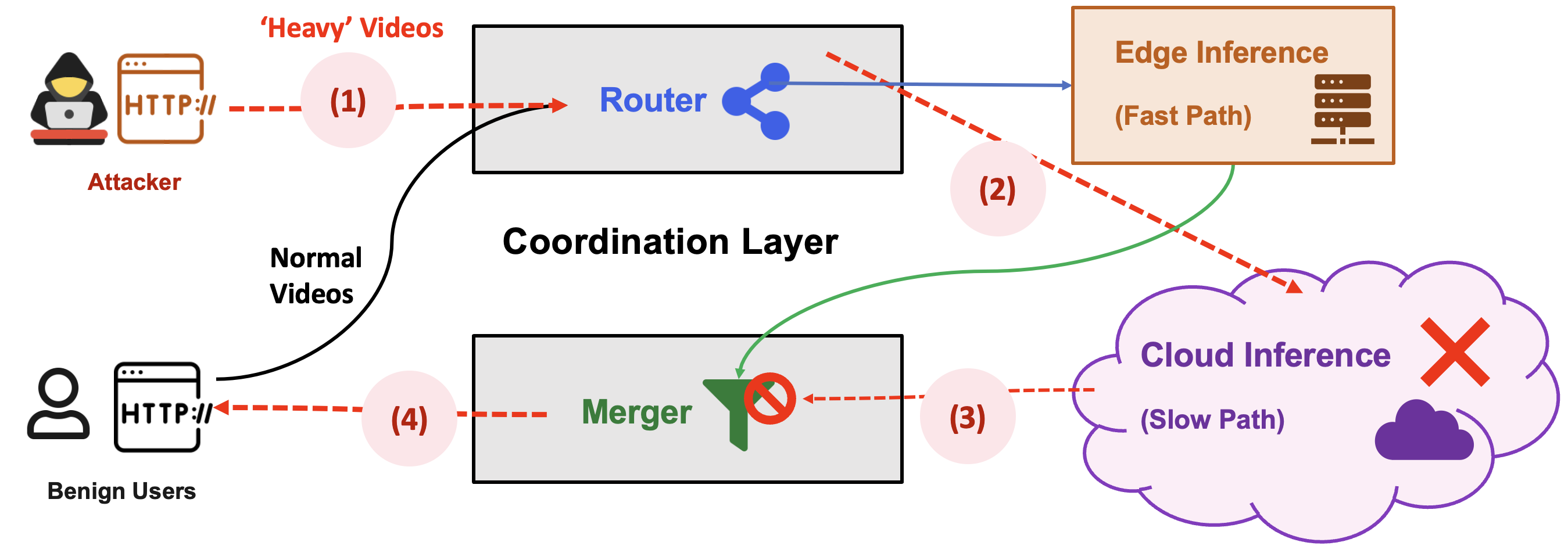}
  \caption{Coordination-layer attack on a slow/fast inference pipeline. (1) The attacker sends a bursty request load that triggers excessive slow-path assistance. (2) The router forwards them to the shared slow path. (3) Contention at shared resources (e.g., scaling and queueing delays) delays results beyond their freshness deadlines, so the merger discards them as stale. (4) Benign users lose the expected benefit of cloud assistance.}
  \label{fig:attack-overview}
\end{figure}

Under this threat model, Figure \ref{fig:attack-overview} shows how an attacker can translate valid network requests into application-level accuracy loss. The attacker submits requests likely to invoke slow-path assistance, for example, video inputs with rapid scene changes or many visible objects (1), which the router forwards to the shared slow path (2). The resulting burst creates contention along the slow path, which may arise from queueing at shared network links or cloud resources (3) (our evaluation focuses on cloud queueing during autoscaling). The merger then discards these predictions as stale, causing benign users to lose the expected accuracy gain from cloud assistance even though the fast path continues to return timely results (4). The system therefore incurs the cost of slow-path computation while realizing only fast-path quality.

We use \emph{accuracy collapse} to denote the loss of the two-system inference pipeline's accuracy gain, potentially reducing quality toward the fast-path baseline.

\section{Evaluations}
\label{sec:eval}

\begin{figure*}[ht]
  \centering
  \begin{subfigure}{0.33\textwidth}
    \centering
    \includegraphics[width=\textwidth]{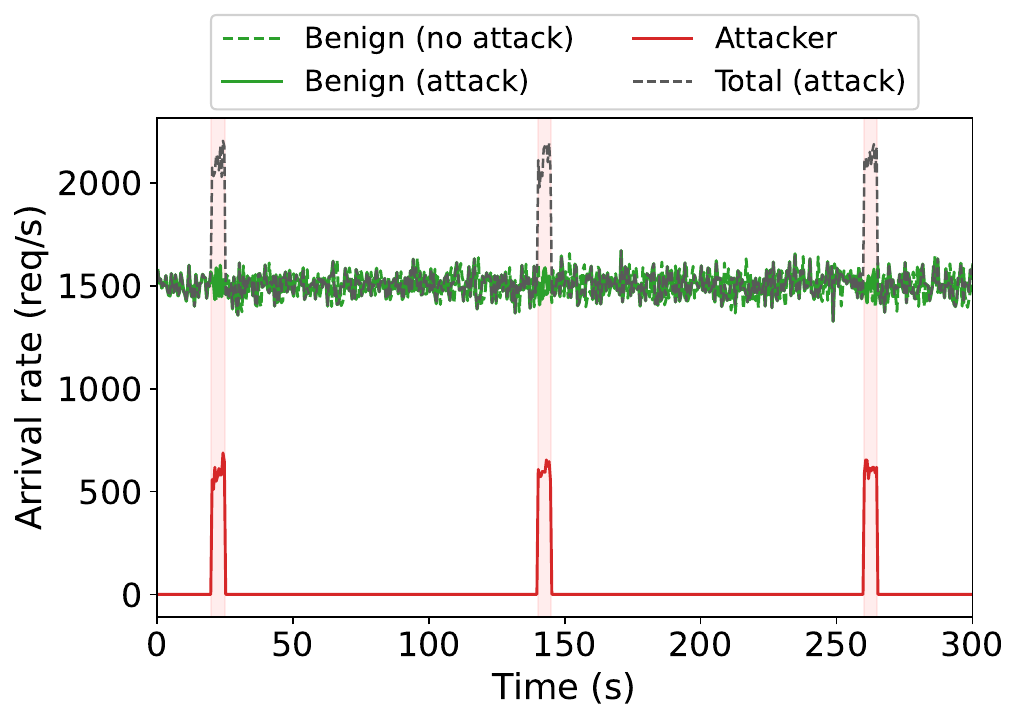}
    \caption{Arrival traffic and attacker bursts.}
   \label{fig:arrivals}
    \end{subfigure}
  \begin{subfigure}{0.33\textwidth}
    \centering
    \includegraphics[width=\textwidth]{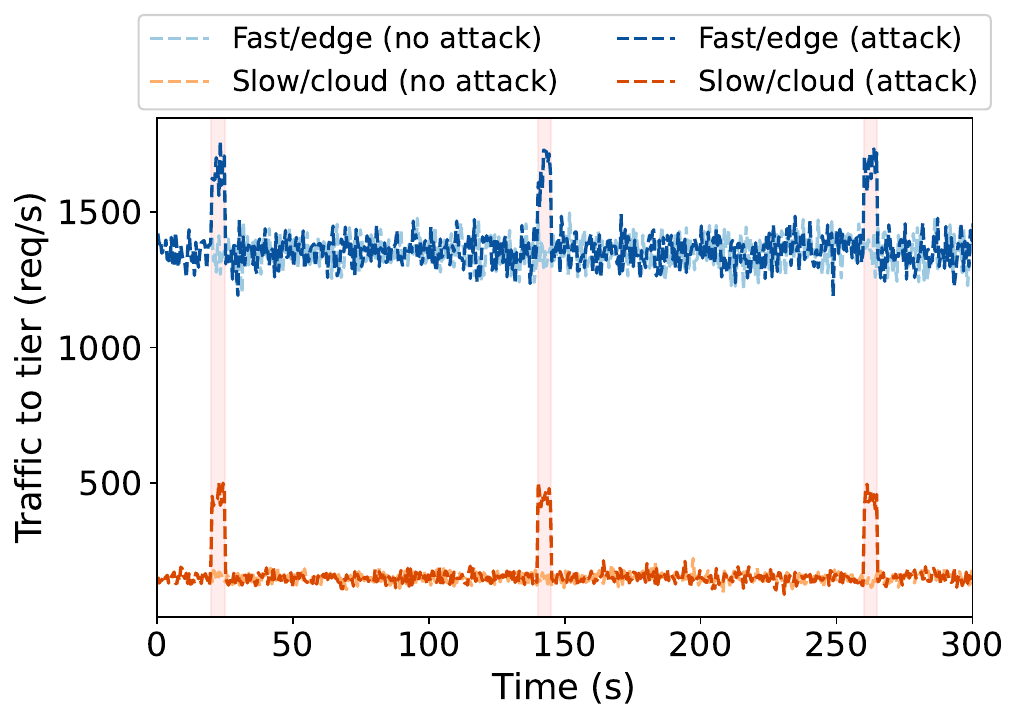}
    \caption{Slow-path and fast-path workload.}
   \label{fig:tiers}
    \end{subfigure}
  \begin{subfigure}{0.33\textwidth}
    \centering
    \includegraphics[width=\textwidth]{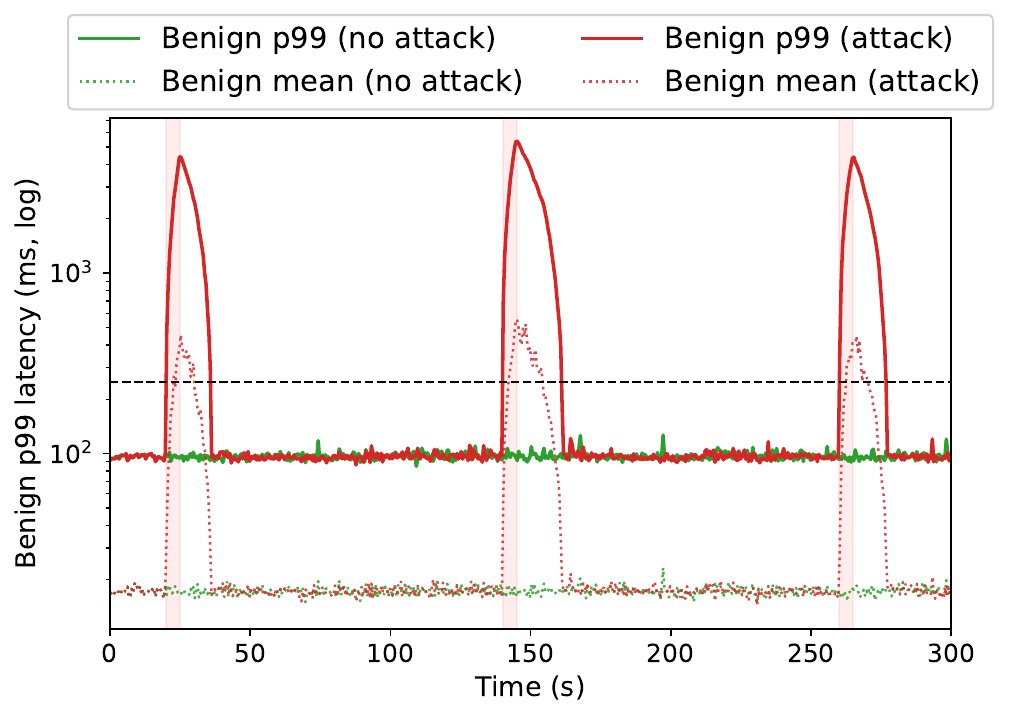}
    \caption{Latency of slow-path requests.}
   \label{fig:latency}
   \end{subfigure}
   \caption{Coordination-layer shaped attack evaluation. Short attacker 'Yo-Yo' style bursts  increase the slow-path (cloud tier) workload, and drive benign slow-path latency well beyond the needed deadline.}
   \label{fig:1a}
\end{figure*}

We evaluate the attack using a two-system inference pipeline, focusing on whether the pipeline preserves the accuracy benefit expected from cloud assistance under attack. This depends directly on tail latency and the application’s freshness SLO, since an otherwise correct cloud prediction is useless if it arrives after its deadline.
We divide the evaluation into two parts. Section \ref{sec:eval-slow-path} shows how bursts of slow-path demand exploit autoscaling lag to increase tail latency. Section \ref{sec:eval-accuracy} then measures how the resulting deadline violations remove cloud assistance and degrade tracking quality over complete videos, targeted intervals, and individual classes.

\subsection{Slow-Path Latency Amplification}
\label{sec:eval-slow-path}

\paragraph{Setup.} We simulate an urban video-analytics system based on the
NSF PAWR COSMOS testbed \cite{raychaudhuri2020challenge} and its
delay-sensitive street-camera analytics settings \cite{ghasemi2025real}.
The simulator uses the assisted edge-cloud architecture from Section \ref{sec:arch}: an edge model provides the fast result, the inference router sends selected requests to a stronger shared cloud model, and a merger discards late cloud assistance. The fast path is a finite multi-server station. The slow path is a multi-server FCFS cloud service whose warm workers are controlled by an autoscaler. 
While attacker-generated load immediately adds to the benign workload, additional workers become available only after cold-start warmup, while idle workers remain provisioned until cooldown ends.

Benign requests represent aggregate cloud invocations from multiple camera streams and arrive as a provisioned Poisson process, with rate $\lambda_b=Nfp$ for $N$ streams at $f$ frames/s and cloud-routing probability $p$. The attacker adds valid requests that trigger more slow-path work and controls only their content and timing. The attack exploits the warmup lag: load rises immediately, but slow-path capacity does not. The attacker spaces bursts so scaled-out workers are reclaimed after cooldown, making each new burst encounter baseline capacity and another warmup delay.

We compare the no-attack baseline against an attacker using  short bursts. We also evaluate a flat high-volume DDoS baseline to separate the effect of bursts 
from sustained load. The ratio between the benign p99 latency, relative to a $250$ ms SLO, and total attacker requests captures the attack potency.

\paragraph{Results.}

Figure~\ref{fig:1a} shows the shaped attack's effect on benign requests. Three short bursts raise total arrival rate (Figure~\ref{fig:arrivals}) and both fast- and slow-path workload (Figure~\ref{fig:tiers}). Without attack, benign p99 latency stays near $100$ ms. During each burst, it rises to several seconds, and mean latency exceeds the $250$ ms SLO (Figure~\ref{fig:latency}). 
Each burst creates contention that delays slow-path completion before additional workers become ready.
The attacker then waits for scaled-out capacity  to be reclaimed, so the next burst again encounters a cold slow path.
Cloud predictions miss their freshness deadline and are discarded, wasting much of the slow-path work and cost. Compared with the flat high-volume baseline, the shaped attack creates more stale-assistance periods and is therefore more harmful to accuracy over time.

\subsection{Model Accuracy Drop}
\label{sec:eval-accuracy}

\begin{figure}[t]
  \centering
  \includegraphics[width=\columnwidth]{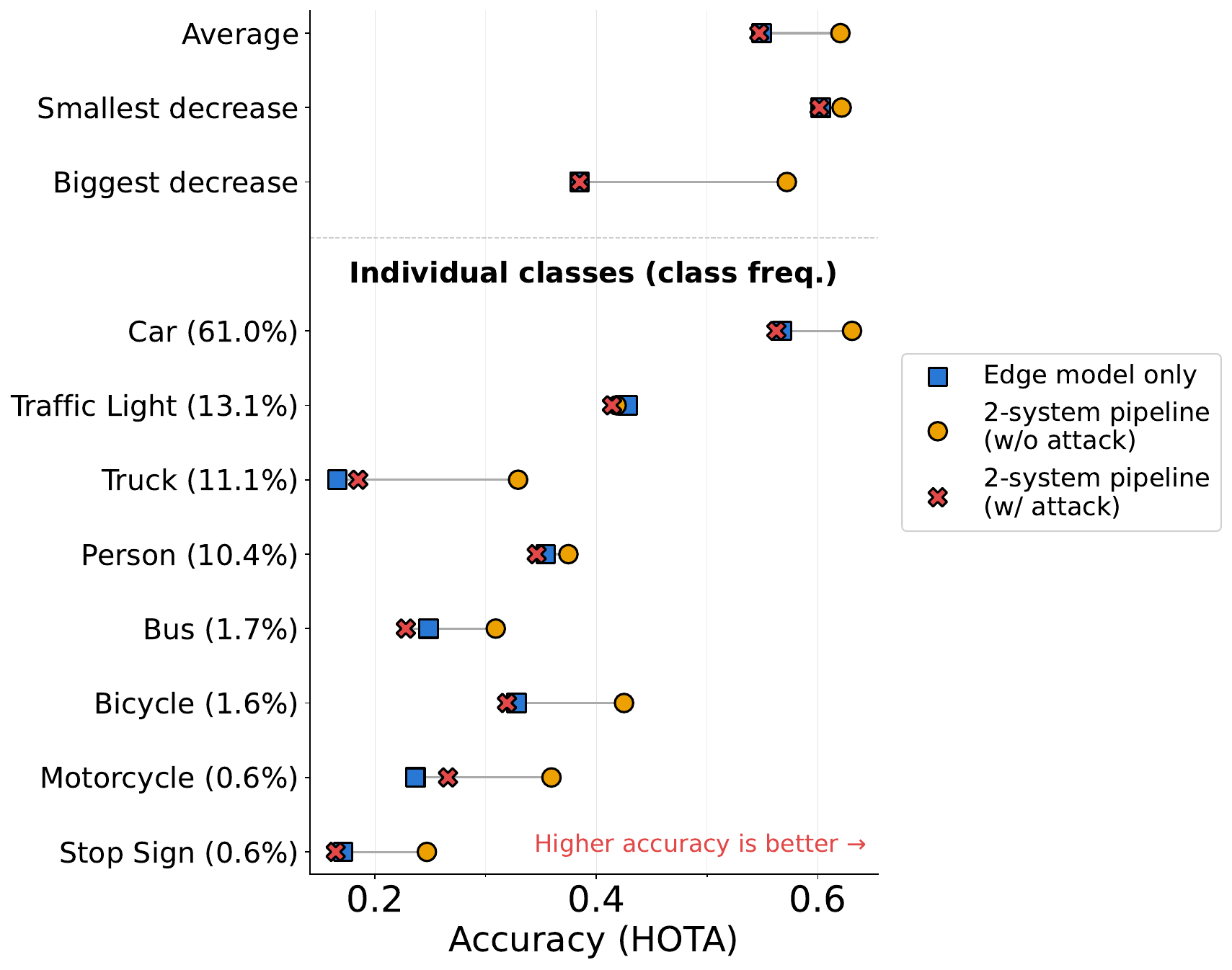}
  \caption{\textbf{Impact of the shaped attack on application accuracy.} The
  two-system architecture normally improves tracking quality over the edge-only
  model. During the attack, delayed cloud predictions
  are discarded as stale, eliminating most of that improvement.}
  \label{fig:dumbbell}
\end{figure}

\paragraph{Setup.} We evaluate whether workload-induced latency translates into
application-level accuracy loss using a real multi-object-tracking workload on
Argoverse-HD~\cite{chang2019argoverse}, an autonomous-driving dataset with
dense annotations designed to evaluate real-time perception in safety-critical
driving settings. The edge tier runs YOLOX-Nano~\cite{ge2021yolox}, while the
cloud tier runs the more accurate YOLOX-X model. The system sends one frame to
the cloud every 5-10 frames at regular intervals. Because cloud detections are
produced only for these sampled frames, we propagate them forward using
Kalman-filter-based object-motion forecasting with camera-motion compensation
(CMC), and then apply the SORT tracker~\cite{bewley2016simple} to associate
objects over time. Consistent with latency budgets used in prior
autonomous-driving perception systems~\cite{gog2022d3,shi2024soar}, cloud
predictions are incorporated only if they return within a 250\,ms deadline;
responses that arrive later are discarded as stale, leaving the edge prediction
as the system output. We report HOTA~\cite{luiten2021hota}, an end-to-end
tracking-quality metric, for the complete videos and the intervals targeted by
the shaped attack. We compare the edge-only baseline against the two-system
pipeline with and without the attack, capturing both its aggregate impact
and its effect during periods of induced slow-path delay.
\\
\\
\textbf{Attack impact on randomly selected targets.}
\Cref{fig:dumbbell}, ``Average'' shows the impact of the attack on accuracy.
Averaged over randomly selected attack placements across the video contents,
the shaped attack reduces tracking quality by 7.0 HOTA points and leaves the
two-system pipeline close to the edge-only baseline, erasing nearly all of the
average benefit provided by cloud assistance.
\\
\textbf{Attack impact in different targeted video content.} The average
results conceal substantial variation across video content. We replay the
identical attack-induced completion-time trace 
over different video intervals
while holding 
load, slow-path latency,
deadline-misses, 
autoscaling, 
and cost
fixed. The results show that operationally equivalent attack executions can
have sharply varied impact on prediction quality: In the least-damaging targets
(\cref{fig:dumbbell}, ``Smallest decrease''), the attack reduces HOTA by only
2.0 points. In the most-damaging targets (\cref{fig:dumbbell}, ``Biggest
decrease''), the same attack reduces HOTA by 18.7 points, more than a
$9\times$ difference in prediction quality harm. Conventional distributed
attack potency metrics assign these executions the same severity and prescribe
the same overprovisioning defense, even though their effects on the application
are radically different. 
The attack's impact therefore depends not only on how
many cloud predictions are late, but on which predictions are lost.
\\
\textbf{Attack impact by class.} The class-specific results expose a second
form of variation. Under attack, the two-system pipeline generally falls toward
edge-only quality, but the fraction of accuracy lost in the attack varies
widely across classes (\Cref{fig:dumbbell}). The most affected class (Truck) loses nearly half of its
accuracy under the attack, and several low-frequency classes, including
bicycles, motorcycles, and stop signs, also lose substantial fractions of their
tracking quality. Aggregate accuracy can therefore hide severe class-specific
regressions. This distinction matters in applications like autonomous driving
because a relatively infrequent object can still be operationally critical to
safe behavior.

\section{Related Work}
\label{sec-relatedwork}

\textbf{From sustained to burst DDoS.}
Traditional DDoS attacks were usually framed as \emph{sustained exhaustion attacks} in which the adversary sent enough traffic, for long enough, to overwhelm a link, server, or service endpoint. In that model, success came from raw volume, so defenses focused on filtering, traffic scrubbing, and rate limiting \cite{peng2007survey,mirkovic2004taxonomy}. 
Over time, services became more distributed, while mechanisms such as retries, overload controllers, and scaling became common \cite{gan2019open,zhou2018overload,deng2024cloud}. Mitigation also improved as 
operators became better at detecting and absorbing persistent floods \cite{peng2007survey,mirkovic2004taxonomy,alcoz2022aggregate}.

Already in 2003, the Shrew attack demonstrated that carefully timed low-rate pulses could exploit protocol reaction dynamics and outrun defenses that rely on observation windows and delayed responses \cite{kuzmanovic2003low}. More recent work similarly showed that reaction time itself becomes part of the attack surface: even with a modest average attack rate, defenses effective in steady state may fail when workloads change faster than controllers can adapt \cite{alcoz2022aggregate,guo2023temporal}.

Further, related burst and low-volume attacks showed that adversaries can target tail rather than mean latency. In web applications and microservices, intermittent traffic can create transient bottlenecks and inflate tail latency while remaining hidden in aggregate utilization \cite{shan2017tail,gu2024sync,gu2024grunt}. Such systems may therefore fail in the tail before the mean.
\\
\textbf{From availability to utility.}
Burst attacks also shifted the objective toward \emph{economic} harm. Rather than continuously denying service, attackers increasingly targeted the cost efficiency of distributed systems. The Yo-Yo attack manipulated cloud elasticity to trigger repeated scale-up and scale-down \cite{sides2015yo,bremler2017ddos}, while denial-of-wallet attacks exploited serverless platforms to drive excessive resource consumption and spending \cite{shen2022gringotts}. In these settings, the service may remain available while incurring disproportionate cost.
\\
\textbf{From utility to accuracy.}
We argue that distributed inference extends this progression one step further. By making slow-path part of the inference process, networking effects can directly shape the accuracy realized by the application. A workload-shaping attacker can exploit shared contention and timing to erase the accuracy benefit of remote assistance while the service remains responsive, making the realized accuracy of distributed inference a target of workload attacks.

\section{Discussion and Future Work}
\label{sec:discussion}

Networking effects such as contention and delay can directly determine the application-level accuracy of emerging distributed inference pipelines. As edge devices become more capable and AI models more demanding, fast/slow two-system architectures are likely to become increasingly common. As we demonstrated, network-level security threats can play a crucial role in determining whether such systems can be trusted in safety-critical domains. We conclude by highlighting the networking questions that arise when attacks on shared infrastructure become attacks on AI accuracy.
\\
\\
\textbf{Worst-Case Coordination.} 
Provisioning enough slow-path capacity to absorb arbitrary bursts seems to be the most direct defense, but is very costly at scale. The broader opportunity is to revisit resilience mechanisms from cloud gateways and shared services, including admission control, rate limiting, traffic isolation, and priority scheduling, through the lens of distributed inference. When can these mechanisms provide strong accuracy guarantees?
\\
\\
\textbf{Other two-system applications.} 
Our evaluation focuses on latency-sensitive perception, but the same abstraction spans robotics, mobile reasoning, routed language models, and other assisted-inference pipelines.
These systems differ in model type, data modality, merger semantics, and the consequences of brief accuracy loss. 
How do these factors determine attack potency, and can common principles predict which applications are most vulnerable?
\\
\\
\textbf{Generalizing Across Coordination Policies.} 
We treated the router and merger abstractly and evaluated an edge-cloud instantiation. Deployed pipelines will span the entire on-device--edge--cloud continuum and employ diverse coordination policies, including selective routing and application-specific merging. How these designs respond to adversarial workloads remains largely unexplored. 
Systematically studying these designs across applications could reveal new attacks and guide the design of resilient distributed inference.

\bibliographystyle{ACM-Reference-Format}
\bibliography{refs_U}

\end{document}